\documentclass{IOS-Book-Article}     
\usepackage{caption}
\usepackage{subcaption}
\usepackage{graphicx}
\usepackage{diagbox}
\usepackage{multirow}
\usepackage{pbox}

\begin{document}
\begin{frontmatter}          
%
\title{Benchmarking the Impact of Noise on Deep Learning-based Classification of Atrial Fibrillation in 12-Lead ECG}
\runningtitle{Impact of Noise on 12-lead ECG Classification Model}

%
\author[A,B]{\fnms{Theresa} \snm{Bender}%
\thanks{Corresponding Author: Theresa Bender, University Medical Center Göttingen, Robert-Koch-Str. 40, 37075 Göttingen, Germany, theresa.bender@med.uni-goettingen.de.}},
\author[A]{\fnms{Philip} \snm{Gemke}},
\author[A]{\fnms{Ennio} \snm{Idrobo-Avila}},
\author[A]{\fnms{Henning} \snm{Dathe}},
\author[A,B]{\fnms{Dagmar} \snm{Krefting}},
\author[A,B]{\fnms{Nicolai} \snm{Spicher}}
\runningauthor{T. Bender et al.}
\address[A]{Department of Medical Informatics, University Medical Center Göttingen, Göttingen, Germany}
\address[B]{DZHK (German Centre for Cardiovascular Research), partner site Göttingen, Göttingen, Germany}

\begin{abstract}
Electrocardiography analysis is widely used in various clinical applications and Deep Learning models for classification tasks are currently in the focus of research. Due to their data-driven character, they bear the potential to handle signal noise efficiently, but its influence on the accuracy of these methods is still unclear. Therefore, we benchmark the influence of four types of noise on the accuracy of a Deep Learning-based method for atrial fibrillation detection in 12-lead electrocardiograms. We use a subset of a publicly available dataset (PTB-XL) and use the metadata provided by human experts regarding noise for assigning a signal quality to each electrocardiogram. Furthermore, we compute a quantitative signal-to-noise ratio for each electrocardiogram. We analyze the accuracy of the Deep Learning model with respect to both metrics and observe that the method can robustly identify atrial fibrillation, even in cases signals are labelled by human experts as being noisy on multiple leads. False positive and false negative rates are slightly worse for data being labelled as noisy. Interestingly, data annotated as showing baseline drift noise results in an accuracy very similar to data without. We conclude that the issue of processing noisy electrocardiography data can be addressed successfully by Deep Learning methods that might not need preprocessing as many conventional methods do.
\end{abstract}


\end{frontmatter}


\section*{Introduction}
Electrocardiograms (ECGs) are recordings of the electrical activity of the heart and are frequently used in emergency and in-patient care. However, different types of noise, either stemming from the patient's behaviour (e.g. motion) or the devices (e.g. power line interference), can be introduced during measurement.
The presence of noise leads to a twofold problem: It impedes detection of anomalies leading to false findings and alarms \cite{Festag.2021} and, if the signal-to-noise ratio (SNR) reaches a certain level, detecting diagnostically-relevant features becomes impossible \cite{Apandi.2020}. 

One class of features with high clinical importance are the so-called ''fiducial points'', i.e. the center, on- and offsets of ECG waves such as the QRS complex and the P-/T-wave. They are used for segmenting heartbeats into meaningful intervals \cite{Spicher.2020} and by doing so allow for arrhythmia detection.
Atrial fibrillation (AF) is the most prevalent arrhythmia which is characterized by uncoordinated electrical impulses in the atrium and might lead to severe cardiovascular issues, such as stroke or heart failure. Analyzing the interval in a heartbeat where a P-wave is expected is crucial for AF classification as its absence indicates a lack of sinoatrial node activity and is thereby a sign for AF \cite{Kreimer.2021}. However, so-called fibrillatory waves might occur, mimicking P-waves, impeding the assessment of sinoatrial node activity.

Many state-of-the-art algorithms for ECG classification are based on extracting semantic features derived from human expert knowledge, such as fiducial points. However, as these algorithms tend to wrong results in case of noise \cite{Kumar.2020}, various 
denoising strategies \cite{Mir.2021} have been proposed.
In contrast, algorithms from the field of deep learning (DL) were explored for ECG classification tasks recently \cite{Ribeiro.2020, Attia.2021}. Instead of semantic features, they are based on agnostic features derived from fully-automatic correlation analysis between input ECGs and output classes in an end-to-end fashion. These models are based on the underlying premise that training and test datasets are stemming from the same distribution, which is often their pitfall in case of dataset shifts (variant devices, users, noise). Although initial studies indicate a better robustness to noise \cite{Venton.2021}, it remains unclear to which extend it affects these models.

Thereby, in this work we benchmark the accuracy of a state-of-the-art pre-trained DL model for 12-lead ECG classification regarding its susceptibility to different types of noise. We use the publicly available PTB-XL dataset which contains annotations for several categories of noise made by human technical experts and compare the model's accuracy w.r.t. type of noise.

\section*{Methods}

We analyze a subset of the PTB-XL dataset containing 12-lead ECGs of 10 second length \cite{Wagner.2020}. It contains all 1,514 ECGs annotated as showing AF (label in PTB-XL: \textit{AFIB}) and we add the first $2,000$ normal ECGs (\textit{NORM}) as healthy controls. For each signal, we use a qualitative and a quantitative method to estimate SNR.

\paragraph{SNR based on annotations} ($\textrm{SNR}_{\textrm{\small a}}$) 
For each ECG we determine the number of noisy leads using the columns \textit{baseline\_drift}, \textit{static\_noise}, \textit{burst\_noise} and \textit{electrodes\_problems} provided in the PTB-XL metadata. In the majority of cases, they contain the name of a single lead (e.g. ``aVL''), multiple leads (``I,aVR'') or ranges (e.g. ``I-III''). Using a custom script, we convert this information to numeric values ranging from $0$ to $12$ for each type of noise. The labels ``alles'' (all) and ``noisy recording'' are converted to $12$. We remove ECGs associated with other labels as they are of a more qualitative nature (e.g. ``leicht'' (light)). In this way, for each signal a qualitative, unit-less, linear SNR measure is computed, ranging from $0$ (no noise reported) to $12*4=48$ (all leads are affected by all types of noise). As shown in Tbl. \ref{tab:my-table}, we use this information to split the dataset in ECGs without (''w/o'') a noise label and ECGs with (''w/'') a noise label.

It has to be underlined that a value of zero does not have to mean that there is no noise, it just reflects that there is a potential for a noise-free ECG. The authors of PTB-XL also indicated that missing annotations in case of artifacts or false annotations in case of noise-free signals might occur. However, they concluded that the metadata bears the potential for ECG quality assessment \cite{Strodthoff.2020}.

\begin{table}[]
\caption{Properties of subset extracted from PTB-XL (left) and  results of DL-based AF classification (right). ECGs are grouped according to annotations: In case there is one or more noise label in the metadata, an ECG is assigned to ''w/'', else to ''w/o''.  FP and FN denote False Positive and False Negative, respectively.}
\label{tab:my-table}
\begin{tabular}{l||c|ccl||c|c}
    Noise Label & AF  & Healthy controls & \  & Noise Label &  DL: FP    & DL: FN \\
    \cline{1-3}
    \cline{5-7}
    w/o   & 1,097       & 1,581  &  & w/o & 0.04 \%     & 3.96 \%     \\
    \cline{1-3}
    \cline{5-7}
    w/ &   417     & 419     &  & w/ & 0.24 \%    & 7.06 \%          
\end{tabular}
\end{table}

\paragraph{Measured SNR} ($\textrm{SNR}_{\textrm{\small m}}$) 
Due to the limitations of the manual annotations and as they are only available for $22\%$ of the PTB-XL database \cite{Strodthoff.2020}, we additionally use a quantitative SNR measure for each signal. We compute the Fourier Transform of the signals as well as the ratio of energies in two frequency bands as proposed in \cite{Haan.2013}. Based on the expected heart rates during AF, we define the ``signal'' frequency band ranging from $40$ to $150$ beats-per-minute ($0.66$ to $2.5$ Hz)  and define the ``noise'' frequency band as $< 40$ and $> 150$  beats-per-minute. By scaling with $10 \log 10$, we arrive at an SNR expressed in logarithmic decibel scale (dB).

\paragraph{DL classification}
ECG data is classified with a pre-trained model by Ribeiro et al. \cite{Ribeiro.2020}. The model is a residual network and was trained on more than two million ECGs that were acquired within a Brazilian telehealth network. It outputs independent probabilities for six abnormalities, but we limit our analysis to AF. We use a threshold defined by the authors\footnote{\url{https://github.com/antonior92/automatic-ecg-diagnosis/blob/master/generate_figures_and_tables.py}, commit 89f929d, line 121}.

\paragraph{Data analysis}
We analyze the subset regarding differences between ECGs with and without noise labels for i) their distribution of $\textrm{SNR}_{\textrm{\small m}}$ and $\textrm{SNR}_{\textrm{\small a}}$ as well as ii) the accuracy of DL classification of each noise category. For ii) we compared the noisy recordings ($\textrm{SNR}_{\textrm{\small a}}>0$) with randomly drawn signals from equally sized control groups ($\textrm{SNR}_{\textrm{\small a}}=0$). 

\section*{Results}

\begin{figure*}[!t]
    \centering
    \begin{subfigure}[b]{0.49\textwidth}
        \centering
        \includegraphics[width=1\textwidth]{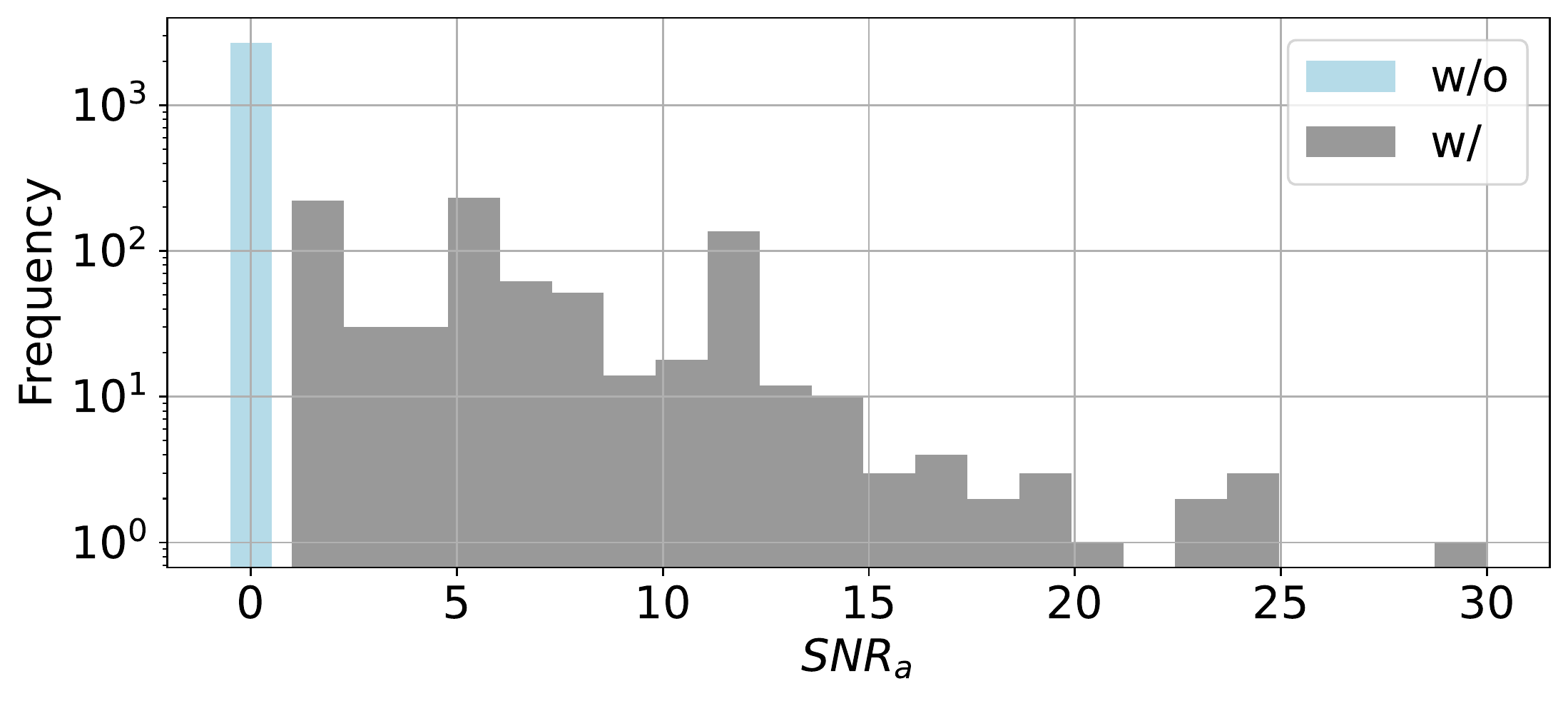}
    	\label{fig-hist-a}
    \end{subfigure}
    \hfill
    \begin{subfigure}[b]{0.49\textwidth}
        \centering
    	\includegraphics[width=1\textwidth]{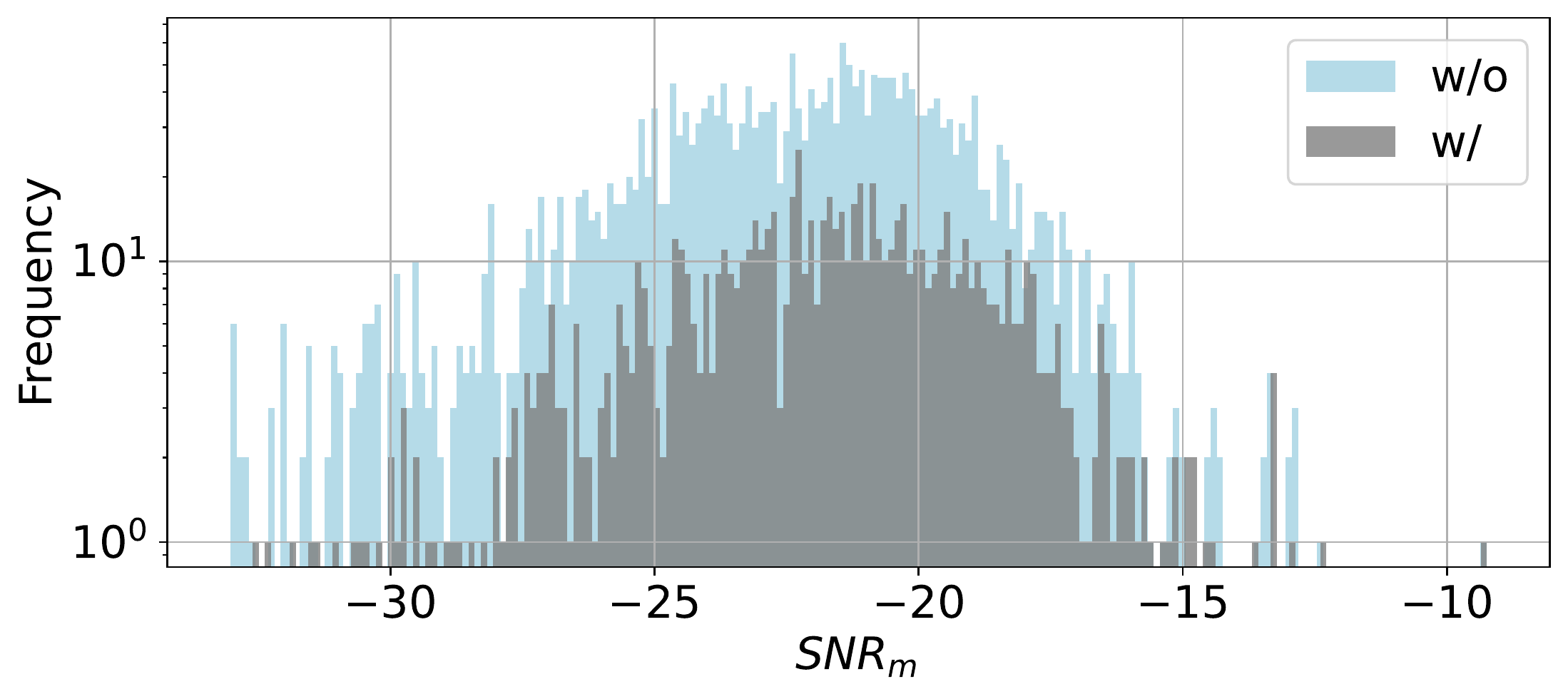}
    	\label{fig-hist-m}
    \end{subfigure}
    \vspace{-0.8cm}
	\caption{Distribution of values of both SNR metrics with (grey) and without (blue) noise labels.}
	\label{fig-steps-single}
\end{figure*}

Fig. \ref{fig-steps-single} shows the distribution of $\textrm{SNR}_{\textrm{\small a}}$ and $\textrm{SNR}_{\textrm{\small m}}$ values on the left and right side. The majority of ECGs with noise labels has less than $15$ with the maximum being $29$. This shows that even in the duration of $10$ seconds, different data quality issues per lead may occur. $\textrm{SNR}_{\textrm{\small m}}$ values are occurring in the range of $[-33.03,-7.78]$ dB with no clear difference between ECGs with and without noise labels.

Tbl. \ref{tab:my-table} (right) shows FP and FN rates of AF classification w.r.t. the existence of noise labels. FP is worsened by $0.2\%$ and FN by $3.1\%$ in case ECGs are annotated with noise labels.
Tbl. \ref{tab:my-table-result} shows the DL accuracy for each type of noise compared to the same number of ECGs but randomly drawn $100$ times from data without noise labels. ECGs with baseline drift or electrode problems are classified more accurately in comparison to random ECG signals without noise annotations, whereas ECGs with annotated burst and static noise reveal worse performance.


\section*{Discussion}
In general, the DL model can robustly classify AF, even in case ECGs are labelled by human experts as having multiple leads influenced by noise. Interestingly, in presence of baseline drift or electrode problems, accuracy is not deteriorated, but within one standard deviation compared to signals without noise labels. As a limitation, it has to be underlined that annotations are non-complete \cite{Strodthoff.2020} and the subset contains only six signals annotated with electrode problems. 

\begin{table}[]
\caption{DL accuracy w.r.t. the four types of noise. The variable $n$ represents the number of signals with the given label (w/). For comparison to signals without a label (w/o), $n$ ECGs are randomly drawn $100$ times and accuracy is given as mean $\pm$ standard deviation.}
\label{tab:my-table-result}
\begin{tabular}{l||c|c|c|c}
\diagbox{Label}{Type} & \pbox{1.8cm}{\centering Baseline Drift \newline ($n = 305$)} & \pbox{1.8cm}{\centering Static Noise \newline ($n = 478$)} & \pbox{1.8cm}{\centering Burst Noise \newline ($n = 156$)} & \pbox{2.5cm}{\centering Electrode Problems ($n = 6$)} \\
    \hline
    \hline
w/o      & $96.8\% \pm 0.9\%$    & $96.8\% \pm 0.7\%$   & $96.9\% \pm 1.3\%$    & $96.3\% \pm 8.0\%$   \\
    \hline
w/    & $97.7\%$       & $94.6\%$    & $94.9\%$      & $100.0\%$                         
\end{tabular}
\end{table}

As the DL model can be assumed as a ''black box'', we can only speculate about the reasons for this behaviour. It could be explained by partial misinterpretation of baseline drift or static noise as P-waves. As we could show in previous work~\cite{Bender.2022}, the DL model was trained such that P-waves and R-peaks have a high relevance, similar to human perception, while numerous other features influence its decision. This multi-factor decision process could be robust to different kinds of noise, but this requires its presence during training.
A shift between training and test datasets is always an issue for DL models. To mitigate this effect is has been suggested to intentionally include noise during training~\cite{Venton.2021}. The model used in this work was trained on $2,000,000$ non-public ECGs. 

However, since the distribution of $\textrm{SNR}_{\small \textrm{m}}$ looks visually similar with or without noise labels, $\textrm{SNR}_{\small \textrm{a}}$ might not be optimal for quality assessment on its own. A ''no noise'' label, explicitly identifying ECGs without data quality issues, and more labels in general would be a valuable addition for future experiments.


\section*{Conclusion}
Results show that the DL model is able to detect AF in 12-lead ECGs with high accuracy, even in the presence of data quality issues according to human experts. We conclude that the difficulty of processing noisy ECGs can be addressed by end-to-end DL models based on agnostic features. In contrast to conventional methods based on semantic features, they might not require preprocessing methods for achieving high accuracy. However, more experiments with larger and more diverse datasets should be the subject of future work.





\end{document}